# Compressibility and pressure-induced disorder in superconducting phase-separated $Cs_{0.72}Fe_{1.57}Se_2$

V. Svitlyk[1], D. Chernyshov[2], A. Bosak[3], E. Pomjakushina[4], A. Krzton-Maziopa[4§],
K. Conder[4], V. Pomjakushin[5], V. Dmitriev[2], G. Garbarino[1], M. Mezouar[1]

[1] ID27 High Pressure Beamline, ESRF, BP220, 38043 Grenoble, France

[2] Swiss–Norwegian Beamlines, ESRF, BP220, F-38043 Grenoble, France

[3] ID28 Inelastic Scattering Beamline, ESRF, BP220, 38043 Grenoble, France

[4] Laboratory for Development and Methods, Paul Scherrer Institute, 5232 Villigen, Switzerland

[§] Present addr.: Faculty of Chemistry, Warsaw University of Technology, 00-664 Warsaw, Poland

[5] Laboratory for Neutron Scattering, Paul Scherrer Institute, 5232 Villigen, Switzerland

E-mail: svitlyk@esrf.fr

**Abstract**

Pressure-dependent diffraction response of the superconducting phase separated $Cs_{0.72}Fe_{1.57}Se_2$ has been studied using synchrotron radiation up to the pressure of 19 GPa. The main and secondary phases of $Cs_{0.72}Fe_{1.57}Se_2$ have been observed in the whole pressure range. The main ordered phase has been found to undergo an order-disorder transition in the Fe-sublattice at least at $P = 11$ GPa with the corresponding kinetics on the order of hours. Contrary to the analogous temperature induced transition, the secondary phase has not been suppressed suggesting that its stability pressure range is higher than 19 GPa or the corresponding transformation kinetics is slower at room temperature. Together with the previously reported pressure-dependent resistivity and magnetic susceptibility measurements, this work indicates that superconductivity in the $A_x Fe_{2-y} Se_2$ ($A$ – alkali metals) phases could be related to the Fe-vacancy ordering in the main phase.

1. **Introduction**

Since its discovery in 2009, the origin of superconductivity in the family of layered Fe-based $A_x$Fe$_{2-y}$Se$_2$ ($A$ – alkali metals) superconductors[1-3] remains unexplained. Furthermore, these compounds exhibit a complex structural behavior. The average structure of the $A_x$Fe$_{2-y}$Se$_2$ compounds corresponds to the ThCr$_2$Si$_2$-type structure (I4/mmm)[4]. At room temperature, the Fe vacancies in $A_x$Fe$_{2-y}$Se$_2$ are ordered resulting in a $\sqrt{5}$x$\sqrt{5}$x1 supercell and the order is lost upon heating [5-7]. In addition, a diffuse scattering commensurate with Bragg reflections from the main phase was observed and related to the correlations in the $A$-deficient sublattice [8].

Series of sharp Bragg peaks and diffuse scattering, especially diffuse rods along c*, not commensurate with the main phase have been observed in the experimental single crystal data. Originally based on X-ray powder diffraction data these features were attributed to an impurity phase resulting from samples' surface degradation and, possibly, to the inhomogeneous distribution of intercalated alkali atoms [5-9]. However, independent diffraction studies [8, 10, 11] proved a regular and consistent nature of the observed features in different samples, thus indicating an intrinsic phase separation in the $A_x$Fe$_{2-y}$Se$_2$ series. The phase separation was directly confirmed by optical, Mössbauer spectroscopies and TEM analysis [12-14]. Our previous diffraction studies[8] on Cs$_x$Fe$_{2-y}$Se$_2$ indicated that the second phase possess a symmetry not higher than monoclinic and that the unit cell is compressed in the $a$-$b$ plane and elongated in the $c$ direction. Diffuse rods along c* indicate the presence of a planar disorder. Monoclinic distortion was also observed in the superconducting Rb$_x$Fe$_{2-y}$Se$_2$ phases[15]. Up to date the detailed structure of the second phase remains unknown, although its average structure can be well described in the ThCr$_2$Si$_2$ I4/mmm model approximation [16].

Pressure-dependent resistivity measurements of the Cs$_{0.83}$Fe$_{1.72}$Se$_2$ phase[5] revealed the suppression of superconductivity at pressures near 8 GPa [17] and the ordering of Fe vacancies was reported to persist up to 12 GPa [5]. In K$_{0.8}$Fe$_{1.7}$Se$_2$ superconductivity was suppressed at 9 GPa [18], however, the resistivity response was slightly different at low pressures compared with K$_{0.6}$Fe$_{1.5}$Se$_2$[19, 20]. For the Rb$_{0.93}$Fe$_{1.70}$Se$_2$ phase, superconductivity disappears near 5.6 GPa [21] similarly to the Rb$_{0.8}$Fe$_{1.6}$Se$_2$ phase [22]. For the latter sample, the $\sqrt{5}$x$\sqrt{5}$x1 supercell indicative of Fe vacancies ordering was reported to persist up to ~15 GPa, in agreement with our previous studies [5]. In addition, in the Rb$_{0.8}$Fe$_{1.6}$Se$_2$ compound, the existence of a third paramagnetic phase around 5.2 GPa was suggested from Mössbauer spectroscopy measurements [22].

Recently, it was shown that after suppression of superconductivity in K$_{0.8}$Fe$_{1.7}$Se$_2$ and K$_{0.8}$Fe$_{1.78}$Se$_2$ at pressures around 9 GPa, superconductivity reemerged near 10.5 GPa with an increased $T_c$ of 48.7 K and disappeared again above 13.2 GPa [23]. Similar behavior was observed for the Tl$_{0.6}$Rb$_{0.4}$Fe$_{1.67}$Se$_2$ system. The reentrant superconductivity phenomenon has been tentatively linked to a pressure-induced phase transition.

Pressure-dependent synchrotron powder diffraction for the same $K_{0.8}Fe_{1.7}Se_2$ and $K_{0.8}Fe_{1.78}Se_2$ samples ruled out the existence of such structural phase transition and confirmed the stability of the tetragonal symmetry of the phases. However, the data were not of sufficient quality to follow the evolution of the Fe vacancies ordering with pressure [23].

Despite the numerous pressure-dependent powder X-ray diffraction studies on the $A_xFe_{2-y}Se_2$ series of compounds even a qualitative description of the structural properties of the second phase has not been reported. In this work, we provide a structural analysis of the pressure-dependant behavior of the second minor phase in the $Cs_xFe_{2-y}Se_2$ system up to a pressure of 3 GPa. We show that the Fe-vacancy superstructure of the main phase is clearly suppressed with pressure and the kinetics of this process is relatively slow (hours time scale).

## 2. Experimental procedure

### 2.1. Single crystal growth and micro X-ray fluorescence spectroscopy

Single crystals of $Cs_{0.72}Fe_{1.57}Se_2$ were grown from the melt using the Bridgman method. The details of the sample preparation are described in the ref [24]. The homogeneity and elemental composition of the cleaved crystal were studied using micro X-ray fluorescence spectroscopy (Orbis Micro-XRF Analyzer, EDAX). Elemental distribution maps for Cs, Fe and Se were collected in vacuum using a white X-ray radiation produced by a Rh-tube (35 kV and 500 µA). The primary X-ray beam was focused down to a spot of 30 µm in diameter. A Ti filter (25 µm thickness) was employed to reject the low energy X-rays. A sample area of ~0.5 cm$^2$ was scanned. Prior to the measurements, elemental calibration was performed using a well characterized standard made of a homogenous mixture of Se, Fe and the corresponding Cs metal carbonate. The obtained composition was $Cs_{0.72}Fe_{1.57}Se_2$ with a ~2% accuracy in the determination of the stoichiometric coefficients.

### 2.2. Pressure-dependent powder diffraction

During the powder diffraction experiment sample handling procedure was similar to the one employed during our previous pressure-dependent studies on the $A_xFe_{2-y}Se_2$ systems[5]. Single crystals of $Cs_{0.72}Fe_{1.57}Se_2$ were finely ground and sealed under an inert argon atmosphere in the glovebox. The resulting sealed powdered sample was opened shortly before loading in the high pressure (HP) diamond anvil cells (DAC).

A first experiment was performed at the Swiss-Norwegian Beamlines at the ESRF, BM01A station in order to study the behaviour of the Fe-vacancies ordering of the main phase with pressure. We used a monochromatic X-ray beam of wavelength $\lambda = 0.6941$ Å and the data collection was performed using a MAR345 detector. A series of X-ray diffraction patterns were collected as a function of pressure up to a maximun pressure of 15 GPa. The sample together with several ruby spheres were loaded in a hole of 0.3 mm in diameter of a stainless steel gasket mounted on a 600 μm diamond. Silicon oil (AP 100) was used as a pressure transmitting medium. The pressure was measured using the ruby luminescence technique [25].

A second high pressure experiment was performed at the High Pressure beamline ID27 at the ESRF. The intense monochromatic X-ray beam of wavelength $\lambda = 0.3738$ Å was generated by a pair of 23 mm period undulators and the data collection was performed using a flat panel Perkin Elmer detector. For this experiment, the sample was loaded in a membrane DAC with helium as a pressure transmitting medium, which preserves excellent hydrostaticity at least up to 50 GPa [26]. The pressure was changed from 0.1 to 19 GPa with a typical step of 0.5 GPa. Similarly to the first experiment, the sample was loaded in a 0.3 mm hole of a stainless steel gasket fixed on a 600 μm diamond and the pressure was measured using the ruby luminescence technique. The powder diffraction data were affected by a pronounced texture which was increasing with pressure (Figure 1, raw 2D powder diffraction data at the pressure of 4.55 GPa is shown as an example).

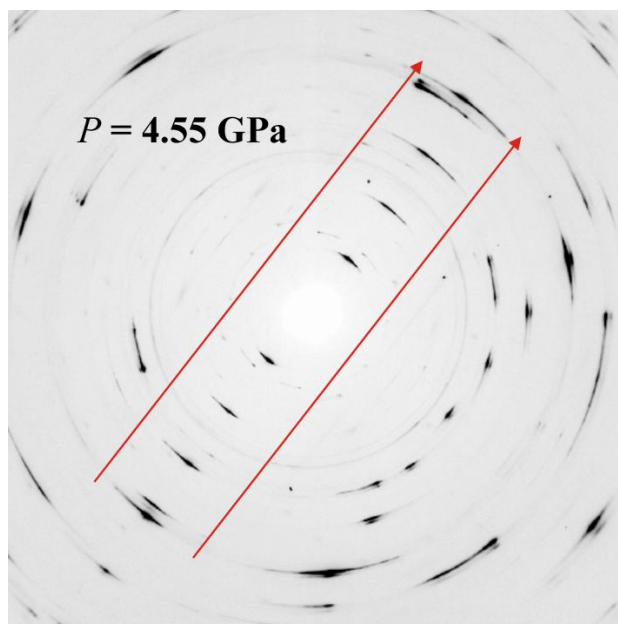

Figure 1. Pressure induced texture in the $Cs_{0.72}Fe_{1.57}Se_2$ sample at $P = 4.55$ GPa. Red arrows indicate the direction of the preferred orientation.

As a result of the sample texturing, a reliable Rietlveld refinement of the structural parameters was not possible. The data were treated using a profile Le Bail fitting method (Fig. 2, refinement of the data at the $P$ = 0.1 (top) and 3 GPa (bottom) is shown), which allowed to obtain the unit cell parameters and unit cell volumes as a function of pressure and, in turn, to calculate the experimental equations of states (EOS). All the data were integrated and processed using the FIT2D software [27, 28]. LeBail fitting of the Powder data were performed using the FullProf software package [29].

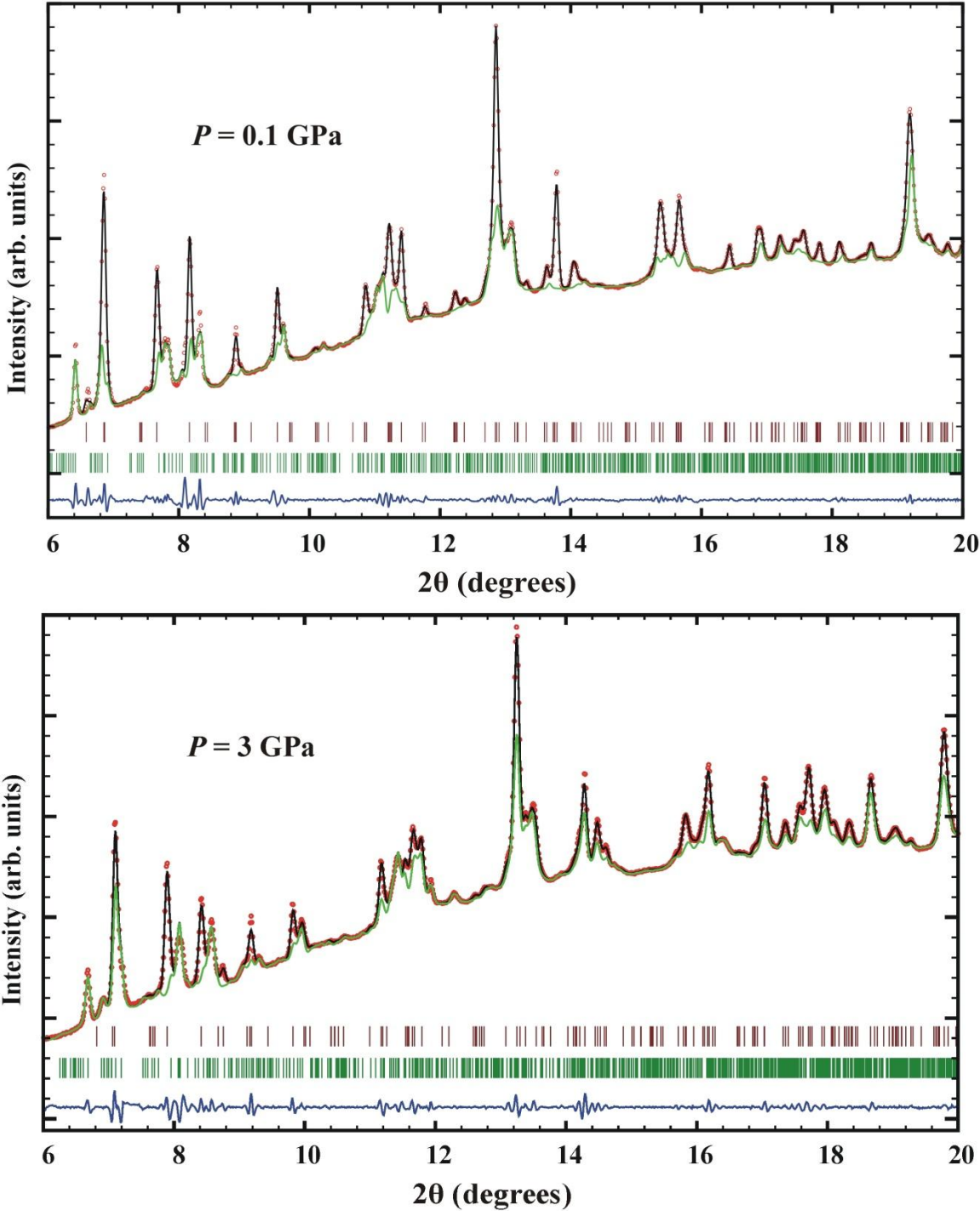

Figure 2. Le Bail profile fitting of the data collected at $P$ = 0.1 GPa (top) and $P$ = 3 GPa (bottom).

Black solid line corresponds to the total calculated contribution from the main and secondary phases; the green solid line corresponds to the calculated contribution from the second phase; the solid blue line is the difference profile; red dots correspond to the experimental profile; the brown and green vertical bars correspond to the Bragg positions of the main and secondary phases, respectively.

## 3. Results and discussion

### 3.4 Pressure suppression of the Fe-vacancy ordering

In our previously published studies on the $A_x Fe_{2-y}Se_2$ (A = Cs, Rb, K) systems as a function of pressure up to 12 GPa [5] no suppression of the Fe-vacancies ordering was observed, i.e. the $I4/m$ symmetry was preserved. Similarly, from the present $P$-dependent studies on $Cs_{0.72}Fe_{1.57}Se_2$ at the SNBL BM01A station, the (110) reflection ($I4/m$ setting) of the main tetragonal phase was not extinct during the initial pressure ramp up to 13.1 GPa. This implies that the ordering of the Fe-vacancies was still preserved at these pressure conditions. Another important aspect resides in the kinetics of a possible order-disorder transition in the Fe sublattice. The total duration of the first experiment including the pressure ramp was about two hours, resulting in a ramp speed of 6.5 GPa/h. At this rate, no order-disorder transition in the Fe sublattice associated with a reduction of the (110) was observed.

To further study the kinetics of this potential structural evolutions under pressure the DAC was left at a constant pressure of 13.1 GPa for a much longer periode of time (t = 18 hours). During that period of time the DAC relaxed to a pressure of 12.4 GPa and the (110) reflection was found to vanish (Fig. 3), thus indicating a transition to the $I4/mmm$ structure with no ordering in the Fe sublattice [4]. The pressure was then released down to to 8.3 GPa and we observed a slight re-appearance of the (110) reflection. Another cycle of pressure increase up to 15.0 GPa erased the peak again, no apparent changes could be observed with a subsequent measurement at 8.2 GPa.

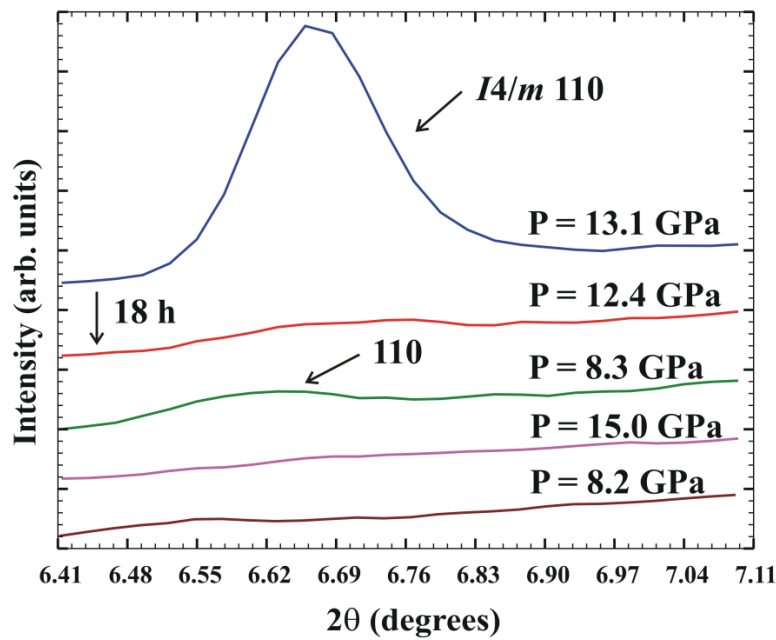

Figure 3. Time- and pressure-dependent evolution of the 110 reflection of the $I4/m$ phase indicative of the Fe-vacancy ordering

In the second series of experiment performed at the high pressure beamline ID27 at the ESRF, we confirmed that the (110) diffraction peak vanished at the $P$ of 11 GPa (Fig 4, a). During this experiment, the pressure was changed in typical steps of about 0.5 GPa and the pressure of 11 GPa was reached in about four hours, which corresponds to a pressure ramp of 2.7 GPa/h.

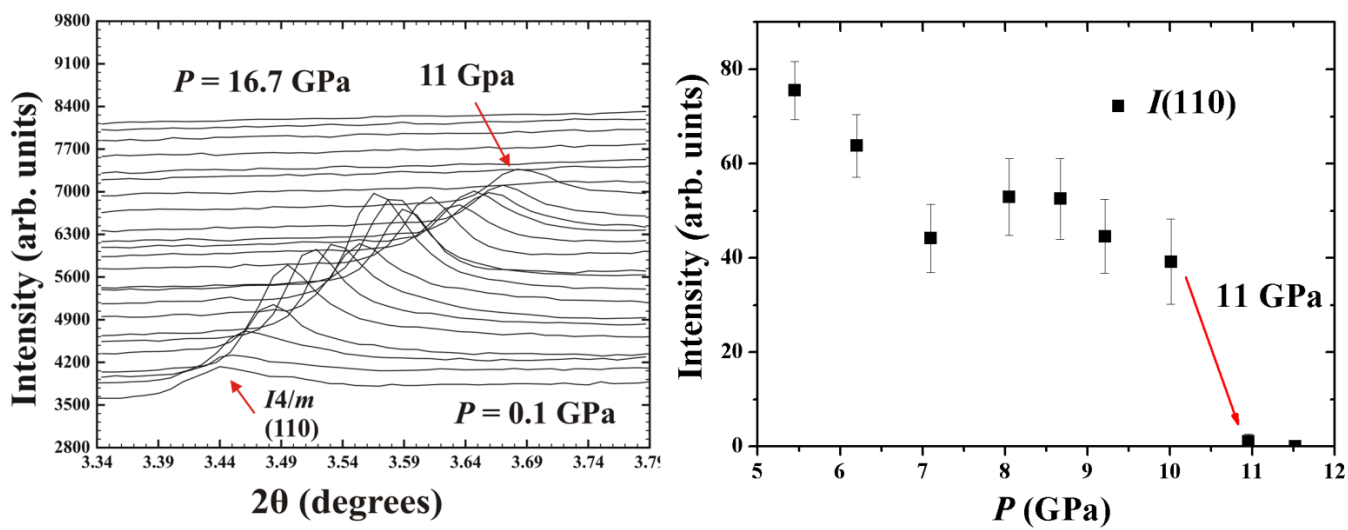

Figure 4. Vanishing of the (110) reflection of the $I4/m$ phase at the pressure of 11 GPa

A sudden step-like disappearance of the (110) reflection (Fig 4, on the right) is consistent with a first-order structural transformation, at least within the resolution of the performed experiment. The first-order transition is also suggested from the initial *P*-dependent run performed at the SNBL BM01A station (Fig. 3). During the first SNBL experiment, the pressure of 13.1 GPa was reached with a rate 2.4 times higher than during the second ID27 experiment. Since the kinetics of the order-disorder transition within the Fe-sublattice is in the order of hours the low-pressure *I*4/*m* phase was "overpressurized" and was seen at the pressure 13.1 GPa. In addition, the behaviour of the order parameter of an analogous *I*4/*m* to *I*4/*mmm* structural transformation observed by us with temperature [4, 5] is also consistent with a first-order transition.

An explicit answer on the order of the observed transition can be obtained from group-theoretical considerations. The structures corresponding to the *I*4/*m* and *I*4/*mmm* symmetries are in a group-subgroup relation[30]. In addition, the transition corresponds to a single $C_1$ irreducible representation (notation of Miller and Love[31]) of the *I*4/*mmm* parent space group[32]. Thus, the first two Landau condition for the second-order phase transitions are fulfilled[33]. However, the Landau expansion of free energy for the above transition contains invariants of a third order[32] which unambiguously indicates that the *I*4/*m* to *I*4/*mmm* structural transformation must be of a first order[33, 34].

### 3.1 Compressibilities, equations of state and pressure-dependant behaviour and of the main and secondary phases in $Cs_{0.72}Fe_{1.57}Se_2$

The pressure evolution of the unit cell volume for the main (*I*4/*m* setting) and secondary phases of $Cs_xFe_{2-y}Se_2$ fitted with a first order Murnaghan equation of state (EoS) (Eq. 1, $V_0$ is the volume at zero pressure, $B_0$ is the bulk modulus and $B_0'$ is the first pressure derivative of the bulk modulus) are shown in Fig. 5. The value of $B_0$ for the main phase above the transition (*I*4/*mmm* symmetry) was fixed to the one obtained from the *F* vs. *f* plot (Fig. 6, see discussion below). The fitted EoS parameters are shown in Table. 1 (the *I*4/*m* setting for the main phase was used).

$$V(P) = V_0(1 + B_0'\frac{P}{B_0})^{-\frac{1}{B_0'}} \quad (1)$$

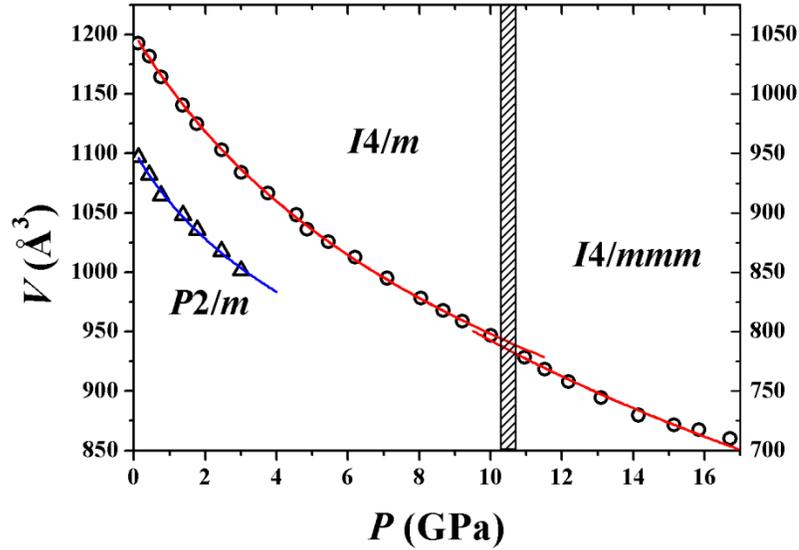

Figure 5. Volume vs. pressure dependences for the main (squares) and secondary (circles) phases in $Cs_{0.72}Fe_{1.57}Se_2$ fitted with a first order Murnaghan equation of state. Right side scale corresponds to the secondary phase

Table 1. Experimental coefficients of the Murnaghan equation of state for the main and secondary phases in $Cs_{0.72}Fe_{1.57}Se_2$

| Symmetry | $P_{range}$, GPa | $V_0$, Å$^3$ | $B_0$, GPa | $B_0'$ |
| --- | --- | --- | --- | --- |
| $I4/m$ | 0.1 – 10.0 | 1199.6(1.2) | 24.1(0.6) | 4.4(0.2) |
| $I4/mmm$ | 10.0 – 16.7 | 1189(6) | 30.5(0.3)* | 2.8(0.2) |
| $P2/m$ | 0.1 – 3.0 | 951.2(3.1) | 19.3(2.7) | 6.3(2.0) |
| $P2/m$ | 0.1 – 3.0 | 949.7(1.8) | 21.6(0.8) | 4.4 (fixed) |

* obtained from the $F$ vs $f$ plot, see discussion below

The analogous $I4/m$-$I4/mmm$ temperature-dependent changes are accompanied by an increase in the unit cell volume [5], which is *a priori* not possible with an application of an external pressure. A presence of a subtle pressure-dependent structural transformations could be tracked using normalized pressure, $F$, vs. Eulerian strain, $f$, dependencies[35-37], where $f = ((V/V_0)^{-2/3}-1)/2$ and $F = P/(3f(1+2f)^{5/2})$.

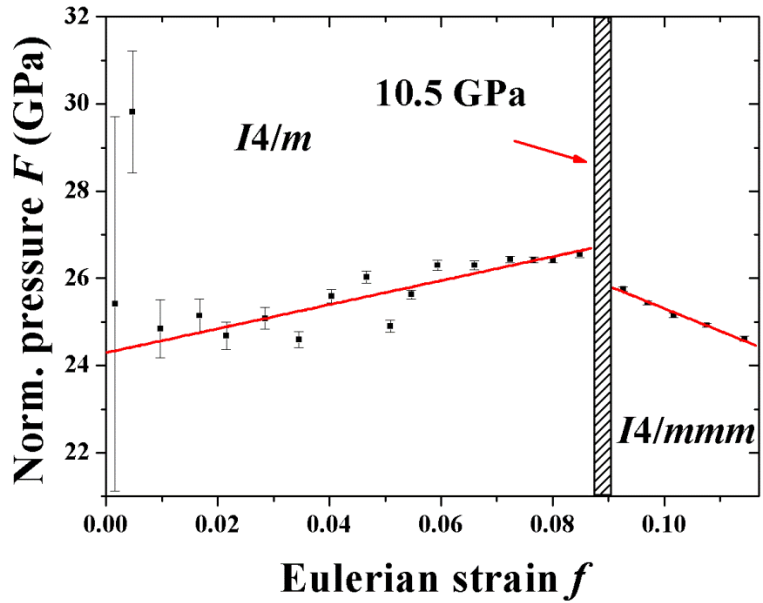

Figure 6. Normalized pressure, $F$, vs. Eulerian strain, $f$, for the main phase showing an anomaly at the $I4/m$ to $I4/mmm$ transition point. Solid red lines correspond to the linear fits for the corresponding regions.

A clear anomaly at the point corresponding to the pressure of 10.5 GPa (Fig. 6) confirms the existence of the $I4/m$ to $I4/mmm$ order-disorder transition. From the $F$ vs. $f$ data bulk moduli are equal to the intersections of the liner fits with the vertical $F$ axe (Fig. 6). The obtained value of $B_0$ for the main phase below the transition point is 24.3(1) GPa and is equal within the error range to the value of 24.1(0.6) GPa (Table 1) obtained from $V$ vs. $P$ the data fitted with a first order Murnaghan EOS. The corresponding $B_0$ value for the secondary phase is equal to 30.5(0.3) GPa (only five first datapoints were included into the fit). From a physical point of view higher values of bulk moduli are expected for high-pressure phases since they are denser and, correspondingly, less compressible.

The secondary phase of $Cs_{0.72}Fe_{1.57}Se_2$ was observed throughout the whole studied pressure range up to 19 GPa. However, the diffraction patterns of the main and, in particular, secondary phases exhibited a gradual degradation (broadening) with pressure (Fig. 7), which could stem from a pressure induced structural amorphization due to a planar morphology of the crystals. As a result, a reliable determination of the cell parameters of the secondary minor phase was not possible at pressures above 3 GPa. Profile fitting of the data collected at $P = 3$ GPa is shown on the figure 2, bottom. The main phase was treated up to the pressure of 16.7 GPa. The refined unit cell parameters for the main and secondary phases are listed in the supporting information for further *ab initio* calculations. Sections of the raw 2D powder patterns illustrating a coexistence of two phases at 0.1 and 16.7 GPa and the $P$-induced profile broadening are presented in Fig. 7.

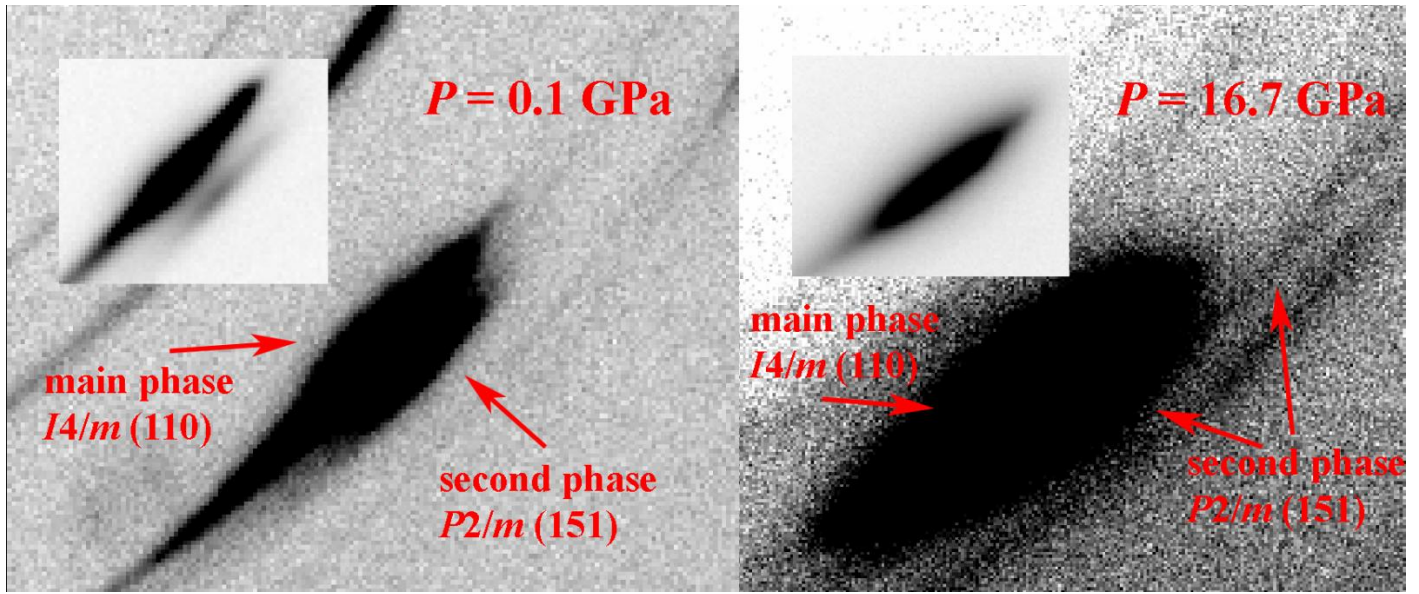

Figure 7. Main and secondary phases of the $Cs_{0.72}Fe_{1.57}Se_2$ sample at the pressures of 0.1 (left) and 16.7 GPa (right). The insets shows the regions with the marked peaks represented with a higher contrast.

No apparent anomalies could be observed in the behaviour of the unit cell parameters for the main and secondary phases (Fig. 8.). However, a clear anomaly can be seen on the $c/a$ ratio of the main phase around the pressure of 11 GPa, which corresponds to the $I4/m$ to $I4/mmm$ transition (Fig. 9). This indicates that the order-disorder transition within the Fe-sublattice is accompanied by subtle anisotropic changes in the unit cell parameters.

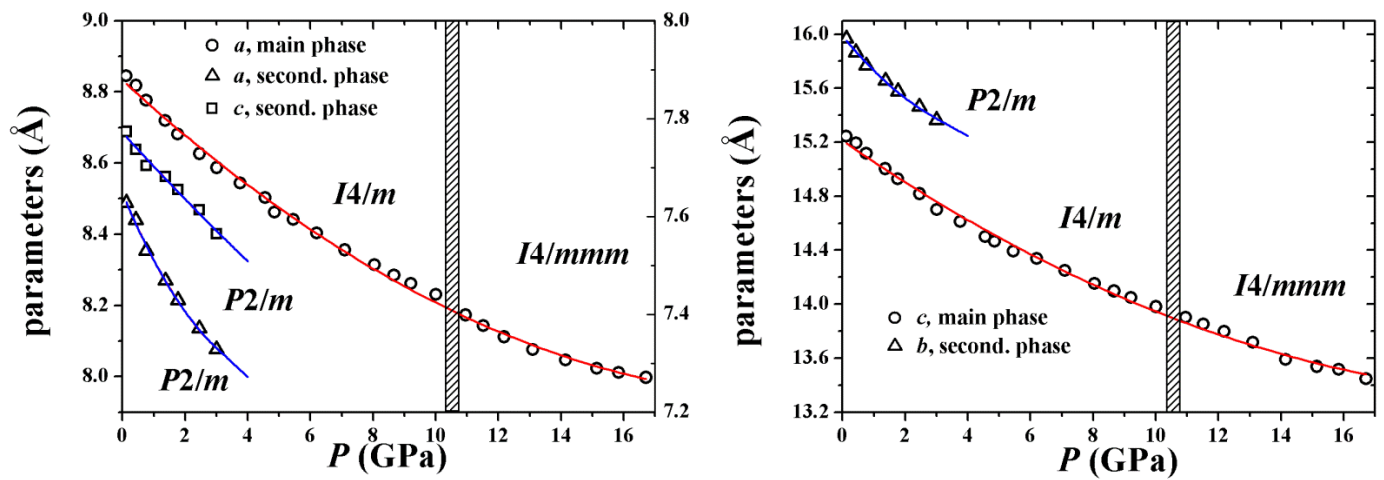

Figure 8. Behaviour of the unit cell parameters for the main and secondary phases. Right-side scale on the left figure corresponds to the secondary phase.

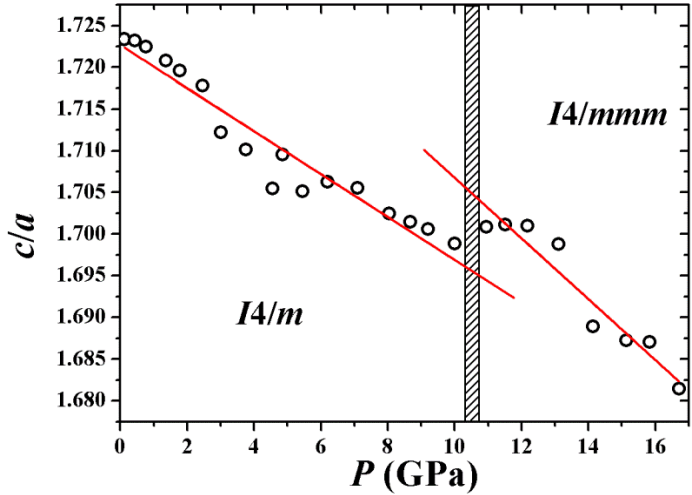

Figure 9. Behaviour of the *c/a* parameters ratio for the main phase indicating an anomaly around 11 GPa.

Interestingly, the jump of the *c/a* ratio during the analogous temperature-dependent *I*4/*m* to *I*4/*mmm* transition reported by in [5] has an opposite sign. The difference in the behaviour of the *c/a* ratio stems from the fact that the temperature-dependent changes are accompanied by a slight increase in the unit cell volume [5]. As was mentioned above, an increase in the unit cell volume with an application of an external pressure is not possible *a priori*. Therefore the structural response during the pressure-induced *I*4/*m* to *I*4/*mmm* transition is more moderate and has an opposite sign.

4. **Conclusions**

Based on pressure-dependent synchrotron powder diffraction experiments, we have characterized the pressure evolution of the main and secondary phases of the phase-separated $Cs_{0.72}Fe_{1.57}Se_2$ superconductor. The (110) Bragg reflection indicative of the Fe-vacancies ordering in the main phase does disappear under pressure manifesting an order-disorder phase transition similar to the one induced by temperature. The critical temperature for vacancy ordering in the $Fe_{2-y}Se_2$ layers should therefore decrease with pressure. Contrary to the temperature-induced transition the kinetics of the analogous pressure-dependant transition is slower and is on the order of hours at room temperature.

Contrary to its temperature evolution, $Cs_{0.72}Fe_{1.57}Se_2$ remains phase-separated at pressures above the order-disorder transition in the main phase, which indicates that the phase separation involves a diffusion of Cs ions that is suppressed or slows down under pressure at room temperature. The different kinetics for the vacancy ordering and phase separation may be potentially used for quenching of various degrees of ordering and separations, in order to manipulate the superconducting fraction of this material.

What phase is superconducting is still an open question. However, the suppression of superconductivity and order in the Fe-sublattice of the main phase with pressure indicates that the main phase in $A_x Fe_{2-y} Se_2$ may be responsible for the observed superconductivity. Definitive conclusions could be made only from diffraction experiments at low temperatures and with identical time scale of the corresponding resistivity measurements.